\documentclass[a4paper]{article}

\usepackage{INTERSPEECH2020}
\usepackage{hyperref}
\usepackage{tabularx}

\title{HooliGAN: Robust, High Quality Neural Vocoding}
\name{Ollie McCarthy\textsuperscript{*}, Zohaib Ahmed}

\address{
  Resemble AI, Toronto, Canada}
\email{ollie@resemble.ai, zohaib@resemble.ai}

\begin{document}

\maketitle
\begin{abstract}
Recent developments in generative models have shown that deep learning combined with traditional digital signal processing (DSP) techniques could successfully generate convincing violin samples \cite{engel2020ddsp}, that source-excitation combined with WaveNet yields high-quality vocoders \cite{wang2019neural, wang2019neural2} and that generative adversarial network (GAN) training can improve naturalness \cite{kumar2019melgan, yamamoto2019parallel}.

By combining the ideas in these models we introduce HooliGAN, a robust vocoder that has state of the art results,  fine-tunes very well to smaller datasets (\textless 30 minutes of speech data) and generates audio at 2.2MHz on GPU and 35kHz on CPU. We also show a simple modification to Tacotron-based models that allows seamless integration with HooliGAN.

Results from our listening tests show the proposed model's ability to consistently output high-quality audio with a variety of datasets, big and small. We  provide samples at the following demo page: \url{https://resemble-ai.github.io/hooligan_demo/}
\end{abstract}


\section{Introduction}
{\let\thefootnote\relax\footnote{{* Corresponding Author}}}
{\let\thefootnote\relax\footnote{{PREPRINT - UNDER REVIEW}}}
Since the introduction of WaveNet \cite{oord2016wavenet}, deep neural network based vocoders have shown to be vastly superior in naturalness compared to traditional parametric vocoders. Unfortunately, the original WaveNet suffers from slow generation performance due to its high complexity and auto-regressive generation. Other works address this issue by reducing parameters and complexity (WaveRNN \cite{kalchbrenner2018efficient}, LPCNet \cite{valin2018lpcnet}, FFTNET \cite{Jin:2018:FAR}) and/or replacing the auto-regressive generation with parallel generation (Parallel WaveNet \cite{oord2017parallel}, WaveGlow \cite{prenger2018waveglow}, Clarinet \cite{ping2018clarinet},MelGAN \cite{kumar2019melgan}, Parallel WaveGAN \cite{yamamoto2019parallel}).

Most of these vocoders consume log mel spectrograms that are predicted by a text to acoustic model such as Tacotron \cite{wang2017tacotron, shen2017natural}. However, if there is sufficient noise in these predicted features entropy increases in the vocoder creating artifacts in the output signal. Phase is also an issue. If a vocoder is trained directly with discrete samples (e.g., with a cross-entropy loss between predicted and ground truth samples) it can result in a characteristic ``smeared'' sound quality. This is because a periodic signal composed of different harmonics can have an infinite amount of variation in its discrete waveform while sounding identical. In this training scenario, the vocoder will be forced to ``solve for phase''.

Differentiable Digital Signal Processing (DDSP) \cite{engel2020ddsp}, while not strictly a vocoder, showed that it is possible to leverage traditional DSP components like oscillators, filters, amplitude envelopes and convolutional reverb to generate convincing violin sounds and that it can be done so without needing to explicitly ``solve for phase''. Meanwhile, the Neural Source Filter (NSF) \cite{wang2019neural, wang2019neural2} family of vocoders show that an F0 (fundamental frequency) driven source-excitation combined with neural filter blocks (i.e., simplified WaveNet blocks) generates outputs with naturalness competitive with vocoders that only take log mel spectrograms as input.

In this work we take the ideas behind DDSP and NSF and combine them into an efficient and robust model, whereby the source excitation is inspired by DDSP and the filtering is inspired by NSF's neural filter blocks. To improve naturalness we also utilise the discriminator module and adversarial training outlined in MelGAN. What we arrive at is a model with impressive sound-quality, inference speed and robustness.

In the next section we review some key ideas from the DDSP, NSF and MelGAN models that we use in this work. In section 3 we outline the HooliGAN model, in section 4 we evaluate the proposed model's naturalness, robustness and performance. We then discuss our findings in section 5.

\section{Background}
\subsection{DDSP}
DDSP \cite{engel2020ddsp} comprises of three main parts, an encoder that takes in log mel spectrograms, a decoder that predicts sequences of parameters that drive an additive oscillator, noise filter coefficients (via spectra predictions), loudness distributions for  the oscillator harmonics and amplitude envelopes for both the waveform and noise. Finally the signal is convolved with a learned impulse response which in effect applies reverberation to the output signal.

While this model excels at generating realistic violin samples, when it comes to modelling speech, we found that it cannot model highly detailed transients on the sample level since the primary waveshaping components, i.e., the filter and envelopes, are operating on the frame level. Also, we found that the sinusoidal oscillator cannot generate convincing speech waveforms on its own.

Our interest in DDSP primarily concerns the additive oscillator and the model's ability to learn time-varying amplitude envelopes.

\subsection{NSF}
NSF \cite{wang2019neural, wang2019neural2} comprises of two excitation sources, namely a fundamental pitch (F0) driven sinusoidal signal and a constant Gaussian noise signal, each of which are then gated by an unvoiced/voiced (UV) switch, filtered by a simplified WaveNet (which they refer to as a neural filter) and passed through a Finite Impulse Response (FIR) filter.

While the sound quality of these vocoders is competitive, we surmise that some of the heavy lifting of the 6 WaveNet stacks could be offloaded to computationally lighter DSP components such as the additive oscillator and loudness envelopes as described in Section 3.2. 

\begin{figure*}[t]
  \centering
  \includegraphics[width=\textwidth]{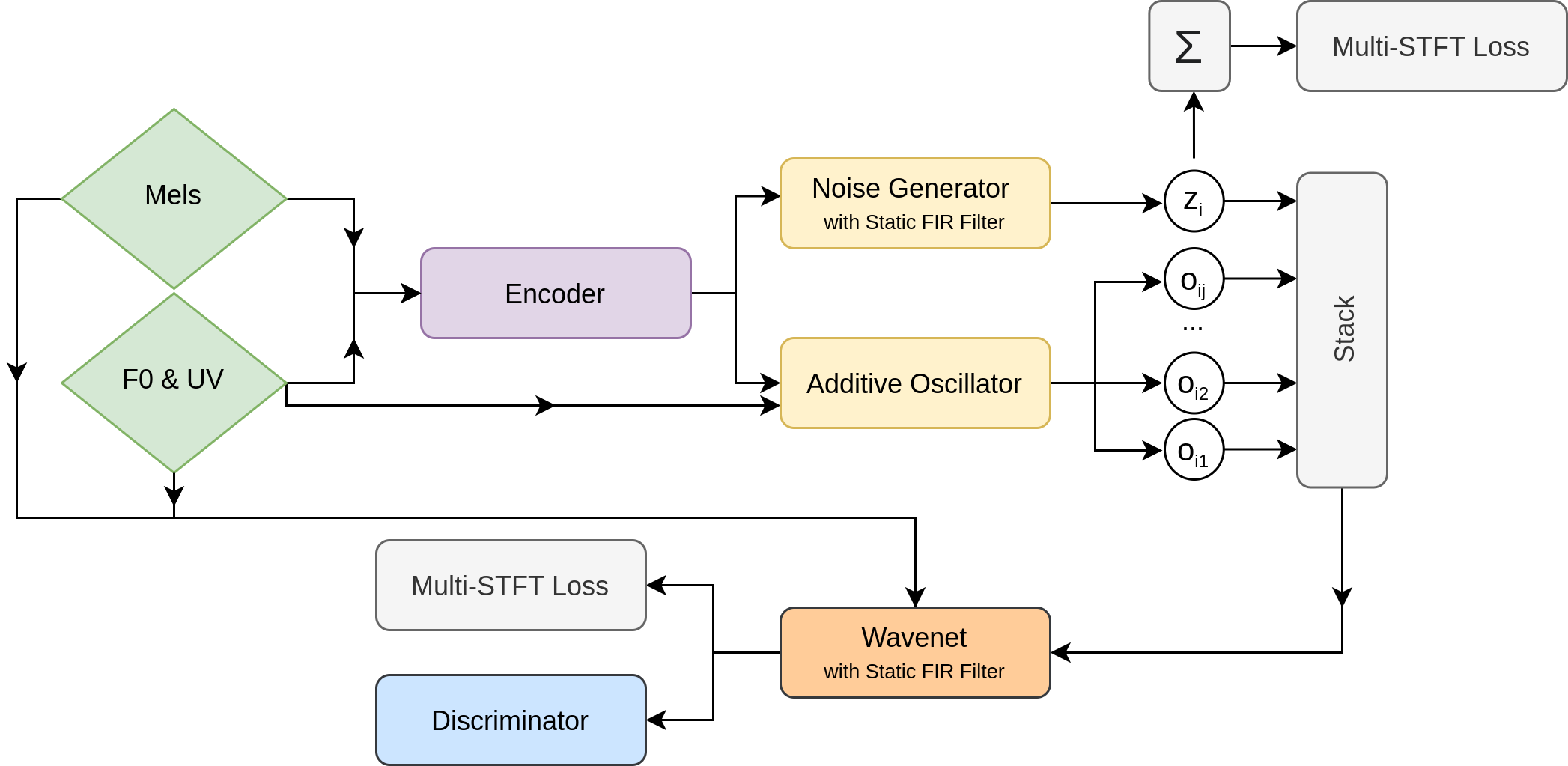}
  \caption{Schematic diagram of HooliGAN.}
  \label{fig:speech_production}
\end{figure*}

\subsection{MelGAN}
MelGAN \cite{kumar2019melgan} and Parallel WaveGAN \cite{yamamoto2019parallel} are the first competitive GAN-based vocoder models.  While Parallel WaveGAN has superior naturalness in its output, MelGAN's discriminator has widely-strided, large kernel size convolutions that are particularly well suited to a phase-agnostic training objective and we incorporate that into our proposed model.

\section{HooliGAN}

In the proposed model, as shown in Figure 1, we take as input log mel spectrograms \(X_{\hat{i}j}\), an F0 pitch sequence \(f_{\hat{i}}\) and a UV voicing sequence \(v_{\hat{i}}\) which we extract from \(f_{\hat{i}}\) with:
\begin{equation}
 v_{\hat{i}} = 
  \begin{cases} 
   1 & \text{if } f_{\hat{i}} > 0 \\
   0       & \text{otherwise}
  \end{cases}
\end{equation}

Note, for all equations \(\hat{i}\) is the time axis on the frame level and \(i\) is the time axis on the sample level while \(j\) will be used interchangeably for the channel/harmonic axes.

\subsection{Encoder}
\(X_{\hat{i}j}\), \(f_{\hat{i}}\) and \(v_{\hat{i}}\) are concatenated together into \(C_{\hat{i}j}\) and put through two 1d convolutional layers with 256 channels, kernel size of 5 and LeakyRELU non-linearity \cite{Maas2013RectifierNI}. Then a fully connected layer outputs a sequence of vectors that is split in two along the channel axis \([H_{osc}, H_{noise}]\) for controlling both the oscillator and noise generator.

\subsection{Additive Oscillator}

The additive oscillator module takes in \(H_{osc}\) from the encoder, transforms it with a fully-connected layer and applies the modified sigmoid non-linearity described in \cite{engel2020ddsp}. We then use linear interpolation to upsample from the frame level to the sample level. Here the parameters are split up into a sequence of distributions \(A_{ij}\) for controlling the loudness of each harmonic and an overall amplitude envelope \(\alpha_i\) for all harmonics.

Before generating the harmonics we create frequency values \(F_{ij}\) for \(k\) harmonics by taking the input F0 sequence \(f_{\hat{i}}\),  upsampling it to the sample-level with linear interpolation to get \(f_{i}\) and multiplying that with the integer sequence:

\begin{equation}
F_{ij} = j f_i\qquad  \forall j \in [1, 2, 3, ..., k].
\end{equation}

Please note that before we upsample \(f{\hat{i}}\), we interpolate across the unvoiced segments in order to avoid glissando artifacts resulting from the quick jumps from 0Hz to the voiced frequency. 

We then create a mask \(M_{ij}\) for all frequency values in \(F_{ij}\) above 3.3kHz. We found that if we did not mask out higher frequencies, the WaveNet would become an identity function early in training and sound quality would not improve with further training. If \(s\) is the sampling rate:
\begin{equation}
 M_{ij} = 
  \begin{cases} 
   0 & \text{if } F_{ij} > (3300/s) \\
   1       & \text{otherwise}
  \end{cases}
\end{equation}

We get the oscillator phase for each time-step and harmonic with the cumulative summation operator along the time axis:
\begin{equation}
  \theta_{ij} = 2\pi \sum_{n=1}^i F_{nj}.
\end{equation}

We then generate all harmonics with:

\begin{equation}
P_{ij} = \alpha_i M_{ij} A_{ij}  \sin(\theta_{ij} + \phi_j),
\end{equation}

\noindent where \(\phi_j\) is randomised with values in [-\(\pi\), \(\pi\)]. Finally we zero out the unvoiced part of the oscillator output \(P_{ij}\) by upsampling \(v_{\hat{i}}\) to \(v_{i}\) with nearest neighbours upsampling and broadcast-multiplying:

\begin{equation}
    O_{ij} = v_i P_{ij}.
\end{equation}

\subsection{Noise Generator}
The noise generator takes in \(H_{noise}\) from the encoder and transforms it with a fully connected layer with modified sigmoid \cite{engel2020ddsp} to sequence \(\beta_{\hat{i}}\). This is upsampled with linear interpolation to the sample level to get \(\beta_i\), the amplitude envelope for the noise. We then get the output with:

\begin{equation}
    z_i = a \beta_i n_i,
\end{equation}

\noindent where \(a\) is a learned parameter initialised with \((2\pi)^{-1}\) and \(n_i \sim \mathcal{N}(0,1)\). We then convolve \(z_i\) with a 257 length impulse response \(h_{noise}\) with learnable parameters:

\begin{equation}
    z_i = z_i * h_{noise}.
\end{equation}

\begin{table}[]
\centering
\caption{MOS test result with a 95\% confidence interval for inverting ground truth acoustic features from LJSpeech. }
\begin{tabularx}{\columnwidth}{X|l}
Model               & MOS\\[1pt]\hline
&\\[-8pt]
MelGAN              & 3.02 ± 0.08\\[1.5pt] 
WaveGlow            & 3.77 ± 0.07\\[1.5pt]
WaveRNN             & 3.77 ± 0.07\\[1.5pt]
\textbf{HooliGAN}   & \textbf{4.07 ± 0.06}\\[1pt]\hline
&\\[-8pt]
Ground Truth        & 4.29 ± 0.06\\
\end{tabularx}
\label{tab:mos1}
\end{table}

\subsection{WaveNet}
We concatenate the stacked harmonics from the oscillator \(O_{ij}\) with shaped noise \(z_i\) to get \(I_{ij}\), the direct input to the WaveNet module. \(C_{\hat{i}j}\) is upsampled with linear interpolation to \(C_{ij}\) and used as side-conditioning in the residual blocks.

We remove the gating function in favour of a simple tanh activation, similar to \cite{wang2019neural, wang2019neural2} and also remove the residual skip collection. However unlike NSF, where the WaveNet channels are reduced to 1 dimension at the output of each stack, we leave the amount of channels constant throughout.  

For the WaveNet hyperparameters we use 3 stacks, each with 10 layers. The convolutional layers have 64 channels and the kernel-size is 5 with a dilation exponent of 2. We also note that this is approximately 2 times more computationally efficient than the NSF WaveNet which have a total 6 stacks. 

Finally we convolve the output of the WaveNet \(w_i\) with a 257 length learned impulse response \(h_{output}\) to get the predicted output:
\begin{equation}
    \hat{y}_i = w_i * h_{output}.
\end{equation}

While this component was originally designed to model reverberation via a long 1-2 second impulse response, early experiments had unwanted echo artifacts. However, with further tweaking we found that a much shorter impulse would help shape the frequency response in the output so we kept this component during development of the model.

\subsection{Training Objectives}
\subsubsection{Spectral Loss}
We use a multi-STFT (Short Time Fourier Transform) loss similar to DDSP \cite{engel2020ddsp} for both the output of the WaveNet \(\hat{y}_i\) and the WaveNet input \(I_{ij}\) where we sum along the stacked axis and add the noise to get \(o_i\) a 1d signal:
\begin{equation}
    o_i = \sum_j I_{ij}.
\end{equation}

\noindent{Then we get the full multi-STFT loss by:}
\begin{equation}
    \mathcal{L}_{stft} = \mathcal{L}_{mag}(y_i, o_i) +
    \mathcal{L}_{mag}(y_i, \hat{y}_i),
\end{equation}
\begin{multline}
    \mathcal{L}_{mag}(y, \hat{y}) = \frac{1}{n}
    \sum_n \biggl( \Vert S_n(y) - S_n(\hat{y}) \Vert_1 \\ +
    \Vert \text{log}S_n(y) - \text{log}S_n(\hat{y})) \Vert_1 \biggl),
\end{multline}

\noindent where \(y_i\) is the ground truth audio and \(S_n\) computes the magnitude of the STFT with FFT sizes \(n \in [2048, 1024, 512, 256, 128, 64]\) and using 75\% overlap.

\subsubsection{Adversarial Losses}
We use similar adversarial training as described in \cite{kumar2019melgan, yamamoto2019parallel} and adapt unofficial open-source code\footnote{https://github.com/kan-bayashi/ParallelWaveGAN} where we use multiple MelGAN discriminators \(D_k\), \( \forall k \in [1, 2, 3]\) of the exact same architecture. To get the generator's adversarial loss \(\mathcal{L}_{adv}\) we use:

\begin{equation}
    \mathcal{L}_{adv} = \frac{1}{k} \sum_k
    \Vert 1 - D_k(\hat{y}_i)\Vert_2.
\end{equation}
The generator's feature matching loss \(\mathcal{L}_{fm}\), where \(l\) denotes each convolutional layer of the discriminator model:
\begin{equation}
    \mathcal{L}_{fm} = \frac{1}{k l}
    \sum_k \sum_l
    \Vert D_k^l(y_i) - D_k^l(\hat{y}_i) \Vert_1.
\end{equation}

\noindent{Our final generator loss \(\mathcal{L}_{G}\), with \(\tau=4\) and \(\lambda=25\) to prevent the multi-STFT loss overpowering the adversarial and feature-matching losses:}
\begin{equation}
    \mathcal{L}_{G} = \mathcal{L}_{stft} + \tau (\mathcal{L}_{adv} + \lambda \mathcal{L}_{fm}) .
\end{equation}

\noindent{The discriminator loss \(\mathcal{L}_{D}\) is calculated with:}
\begin{equation}
    \mathcal{L}_D = \frac{1}{k} \sum_k
    \biggl (
    \Vert 1 - D_k(y_i) \Vert_2 + 
    \Vert D_k(\hat{y}_i) \Vert_2
    \biggl ) .
\end{equation}

\begin{table}[]
\centering
\caption{MOS test result with a 95\% confidence interval for inverting acoustic features predicted by Tacotron2 trained on small datasets.}
\begin{tabularx}{\columnwidth}{l|X|l}
Dataset & Vocoder Model         & MOS\\[1pt]\hline
&&\\[-8pt]
SpeakerA & WaveRNN              & 4.10 ± 0.07\\[1.5pt] 
SpeakerA & \textbf{HooliGAN}    & \textbf{4.49 ± 0.05}\\[1.5pt]\hline
&&\\[-8pt]
SpeakerB & WaveRNN              & 3.18 ± 0.09\\[1.5pt]
SpeakerB & \textbf{HooliGAN}    & \textbf{4.05 ± 0.06}\\[1pt]\hline
\end{tabularx}
\label{tab:mos2}
\end{table}

\subsubsection{Training Schedule}
First we pretrain the generator with only the multi-STFT loss \(\mathcal{L}_{stft}\) for 100k steps similar to \cite{yamamoto2019parallel} after which we switch to the full adversarial loss (\(\mathcal{L}_{G}\) and \(\mathcal{L}_{D}\)). We use the RAdam optimiser \cite{liu2019variance} with a fixed learning rate throughout of \(10^{-4}\) for the generator and \(5 \times 10^{-5}\) for the discriminator with \(\epsilon = 10^{-6}\) and no weight-decay for both optimisers.

We train with a batch size of 16 with \(y_i\) having a duration 11,000 samples. The mel-spectrograms have 80 frequency bins, a hop-size of 12.5ms, a window of 50ms and FFT-size of 2048. We alter the F0 frames from Parselmouth/Praat such that they are centered like Librosa's spectrograms and have an equal hop-size. We use a sampling rate of 22.05kHz. A single Nvidia 2080TI-RTX card is utilised for all experiments and we train for 500k - 1.5M steps. 

\subsection{Tacotron2 Modifications}

Our text to acoustic model is Tacotron2 \cite{shen2017natural} which we modify as follows:
\begin{enumerate}
  \item During training, we predict an F0 value along with each mel-spectrogram frame and re-scale the F0 to be in range [0, 1] by dividing by the maximum frequency parameter in the F0 estimation algorithm. During inference, we re-scale back to hertz and divide by the sample-rate to get the correct frequency value for the oscillator.
  \item The predicted F0 bypasses the prenet and instead is concatenated to the prenet output and then reshaped with a fully connected layer before entering the decoder RNN.
  \item the training objective is altered to become:
  \begin{equation}
      \mathcal{L}_{tts} = \Vert S_{\hat{i}j} - \hat{S}_{\hat{i}j} \Vert_2 + \kappa \Vert f_{\hat{i}} - \hat{f}_{\hat{i}} \Vert_1 .
  \end{equation}
\noindent where \(S_{\hat{i}j}\) is the mel spectrogram and \(\kappa = 2\).
\end{enumerate}

\begin{table}[]
\centering
\caption{MOS test result with a 95\% confidence interval for inverting Tacotron2 predicted features that was trained the LJSpeech Dataset.}
\begin{tabularx}{\columnwidth}{X|l}
Model               & MOS\\[1pt]\hline
&\\[-8pt]
WaveRNN             & 3.65 ± 0.08\\[1.5pt] 

\textbf{HooliGAN}   & \textbf{4.18 ± 0.07}\\[1pt]\hline
&\\[-8pt]
Ground Truth        & 4.28 ± 0.06\\ 
\end{tabularx}
\label{tab:mos3}
\end{table}

\section{Experiments}
\subsection{Datasets}
We use four English language datasets in total for testing sound-quality:
\begin{enumerate}
    \item \textbf{LJSpeech \cite{ljspeech17}}: a 24 hour single-speaker female dataset featuring Linda Johnson from LibriVox with 13,100 transcribed utterances.
    \item \textbf{VCTK \cite{vctk}}: a multi-speaker dataset with 109 unique speakers. Each speaker reads approximately 400 sentences. We downsample from 48kHz to 22.05kHz for our experiments.
    \item \textbf{SpeakerA}: a 30 minute proprietary single-speaker female dataset with 582 utterances.
    \item \textbf{SpeakerB}: a 30 minute proprietary single-speaker male dataset with 500 utterances.
\end{enumerate}

\subsection{MOS Design}

For each experiment we recruit 30 workers from Amazon Mechanical Turk (AMT) for a Mean Opinion Score (MOS) study. We require that the workers have AMT's high-performance ``Master'' status and live in English-speaking countries. We evaluate each model in each experiment with 20 samples. In order to avoid the ``louder sounds better'' perceptual bias we normalise all audio to -23 LUFS using the EBU R128 loudness standard\footnote{https://tech.ebu.ch/docs/r/r128-2014.pdf}. 

\subsection{Experiment Setup}

To test the proposed model's sound-quality we design four experiments. 

\subsubsection{Analysis / Synthesis}
We compare the ability to invert ground-truth acoustic features of the proposed model against WaveRNN, MelGAN and WaveGlow. For MelGAN and WaveGlow we use the pretrained models released by Descript and Nvidia on Pytorch Hub\footnote{https://pytorch.org/hub/}. For WaveRNN we use the pretrained Mixture of Logistics (MOL) model available in our github repository\footnote{https://github.com/fatchord/WaveRNN}. All pretrained models are trained on the LJSpeech dataset \cite{ljspeech17}. Since we cannot fully control the train/validation/test split of all these pretrained models we need to ensure that the evaluation data we use was not seen by the models during training. To this end, we gather some recordings of Linda Johnson that were published \emph{after} the release of the original LJSpeech dataset\footnote{https://librivox.org/the-great-events-by-famous-historians-volume-3-by-charles-f-horne/}. We note that these newer recordings have an almost identical recording quality to those in the original dataset.

In \autoref{tab:mos1} we see that HooliGAN achieves a leading MOS score of 4.07. While the output is clear and high-quality, we do notice that the transients can sometimes be too short, with a click-like quality. We also note that both WaveRNN and WaveGlow have a ``smeared'' characteristic in their outputs, most likely from those models being trained directly on discrete waveform. While we surmise that MelGAN performed poorly mainly because its generator network is under-powered and its receptive field too small. 

\begin{table}[]
\centering
\caption{Inference speeds for all models in our experiments. Default Pytorch settings for Intel(R) Core(TM) i9-7920X CPU @ 2.90GHz and an Nvidia 2080Ti-RTX GPU}
\begin{tabularx}{\columnwidth}{X|l|l|l}
Model               &Parameters     & CPU       & GPU       \\[1pt]\hline
&&&\\[-8pt]
WaveGlow            &87.9M          & 6kHz      & 155kHz    \\[1.5pt]
WaveRNN             &4.5M           & 20kHz     & 43kHz     \\[1.5pt]
\textbf{HooliGAN}   &1.3M           & 35kHz     & 2.2MHz    \\[1.5pt]
MelGAN              &4.3M           & 272kHz    & 3.6MHz    \\[1.5pt]

\end{tabularx}
\label{tab:speed}
\end{table}

\subsubsection{Text-to-Speech Vocoding}

In the second experiment, we test the ability to invert acoustic features predicted by Tacotron2 trained on LJSpeech. We pick sentences from the same evaluation data in Section 4.3.1.
\autoref{tab:mos3} summarises the results with HooliGAN outperforming WaveRNN by a wide margin. We also note that the HooliGAN MOS is quite close to that of ground truth in this experiment.

\subsubsection{Finetuning on Small Datasets}

In the third experiment, we compare the combination of Tacotron2 with the proposed model and WaveRNN when finetuning on datasets with only 30 minutes of data. We pretrain Tacotron2 and the vocoder models with VCTK and finetune afterwards, picking the best performing checkpoints for all models.

In \autoref{tab:mos2} we see again that HooliGAN outperforms WaveRNN. While there is an improvement for both speakers, we note that the improvement is larger for the male speaker. We conclude that having an explicit pitch signal feeding the vocoder is more important for male voices than female as the harmonics tend to be too compressed in mel spectrograms from male voices.

\subsubsection{Inference Speed}

Finally, we test the inference speed of all models in this paper for both CPU and GPU with standard, non-optimised Pytorch\cite{NEURIPS2019_9015} code. In \autoref{tab:speed} we see that MelGAN clearly out-performs all models. However HooliGAN still performs respectably, and while it has less parameters than MelGAN, the deeply-stacked nature of WaveNet limits overall speed. WaveRNN is slowed down by its auto-regressive generation and WaveGlow's theoretically fast parallel generation is limited by high complexity from the large parameter count.

\section{Conclusions and Future Work}

As we can see from the outlined experiments, HooliGAN outperforms all tested models by a large margin in a variety of testing scenarios. We conclude that the source excitation method combined with traditional DSP techniques not only reduces the complexity of the model, but also improves the overall sound quality. In future work we will explore ways to better model transients, investigate other methods of source excitation to further reduce complexity, explicitly model background noise with DSP components and raise the sampling rate to CD-quality 44.1kHz.  

\section{Acknowledgements}

We would like to thank Jeremey Hsu, Corentin Jemine, John Meade, Zihan Jin, Zak Semenov, Aditya Tirumala Bukkapatnam, Haris Khan, Tedi Papajorgji and Saqib Muhammad from Resemble AI for their feedback and support. 

\bibliographystyle{IEEEtran}

\bibliography{mybib}

\end{document}